\documentclass[preprint]{ptephy_v1}%%%%%% to generate preprint number with ptep logo

\preprintnumber{XXXX-XXXX} %%% %%% Insert preprint number here
\usepackage{hyperref}
\begin{document}

\title{Medium effects of charged-hadron production in  $p+Pb$  and $Pb+Pb$ collisions at LHC energies using modified Tsallis distribution}

%%%% To generate auto affiliation numbers please use \author{}\affil{} command

\author{Kapil Saraswat}
\affil{Institute of Physics, Academia Sinica, Taipei 11529, Taiwan
\email{drkapilsaraswat@zohomail.com}}

\author{Prashanta Kumar Khandai}
\affil{Department of Physics, Ewing Christian College, Prayagraj 211003, India
\email{pkkhandai@gmail.com}} 

\author{Deependra Singh Rawat}
\affil{School of Allied Sciences (Physics), Graphic Era Hill University, 
Bhimtal Campus, Sattal Road, Nainital 263132 India 
\email{dsrawatphysics@gmail.com}}

\author{Venktesh Singh}
\affil{Department of Physics, School of Physical \& Chemical Science, 
Central University of South Bihar, Gaya 428236, India
\email{venktesh@cusb.ac.in}}

%\author{Insert last author name here\thanks{These authors contributed equally to this work}}
%\affil{Insert last author address here}

%%% To include the collaborator name... Please use the command "\collaborator"
%%% For example: \collaborator{ATLAS Collaboration}

\begin{abstract}
  The transverse momentum ($p_T$) spectra of charged hadrons in $p+p$, $p+Pb$ and $Pb+Pb$ collisions 
  at $\sqrt {s_{NN}} = 5.02$ TeV are presented here within the rapidity 
  range of $-2.5<y<2.0$. We study the medium effects, which is produced by heavy ion collisions, 
  on the behaviour of charged hadrons, by using a phenomenological fit function. These effects are attributed to two main factors: 
  the transverse collective flow  and the the energy loss of charged hadrons due to multiple scatterings. We observe the transverse
  collective flow at low and intermediate $p_T$ region and the energy loss at high $p_T$ region. 
  Here we take all the published data from the ATLAS collaboration.\\  
\end{abstract}

\subjectindex{Charged hadron spectra, Quark Gluon Plasma, Collective Flow, Energy loss}

\maketitle

\section{Introduction}
The RHIC (Relativistic Heavy Ion Collider) \cite {RHIC1, RHIC2} and the LHC (Large Hadron Collider)  \cite{LHC} are 
designed to study an elusive state of matter known as Quark Gluon Plasma (QGP) \cite{Shuryak}. 
The QGP is a thermalized state of matter where quarks and gluons are deconfined and free to move independently.
Scientists believe that such a state would have existed in the early Universe shortly after the Big Bang \cite{Olive,Schwarz}.
There are many heavy-ion collisions performed at RHIC such as $Au+Au$, $Cu+Cu$, $Cu+Au$ etc
and many at LHC as $Pb+Pb$, $Xe+Xe$ etc. The $p+p$ collisions are taken as a baseline for collisions involving heavy ions \cite{PPPROD} .
The particles produced in both $p+p$ and heavy ion ($A+A$) collisions are the results of multiple
scatterings among partons. The distribution of $p_T$ spectra in $p+p$
collisions provides insights into the state of the matter when particles stop interacting at freeze-out state.
Recent high-multiplicity  $p+p$ collisions data from the LHC \cite{HighMul, HighMul2} predicts the formation of a quark-gluon plasma-like medium.\\

In high energy collisions between heavy ions (like lead or gold), the momentum distribution of emitted hadrons
exhibits extra effects that happen after the initial collision. These effects include the collective flow \cite{Collective_flow}
of hadrons, caused by the expansion of a hot \cite{Recom_Model1, Recom_Model2}, dense matter formed in the collision and
jet quenching  \cite{Jet_Quenching}, where high energy sprays of particles lose energy as they travel through the dense environment.\\
In $p+p$ collisions, researchers analyze the $p_{T}$ spectra of hadrons using a Tsallis distribution ~\cite{Tsallis:1987eu, Biro:2008hz}
characterized by two parameters $T$ and $q$  \cite{PPG099}. While the $T$ parameter corresponds to the kinetic freeze-out temperature
in heavy ion collisions, where particles stop interacting elastically, its interpretation in $p+p$ collisions is less clear
compared to different types of collisions. The parameter $q$, known as the nonextensive parameter indicates how much the system
deviates from complete thermalization  \cite{q_Tsallis}. It captures variations in the system's temperature
\cite{IJMPA_Khandai,  Wong:2012zr, Wong:2013sca}. The Tsallis distribution, which models systems close to thermal equilibrium,
resembles Hagedorn's power law, which is employed in hard scattering processes in QCD ~\cite{Hagedorn:1983wk, Blankenbecler:1974tm}.\\

This work focuses on the $p_{T}$ distribution of charged hadrons in $p+p$, $p+Pb$ and $Pb+Pb$ collisions at $\sqrt {s_{NN}} = 5.02$ TeV
from the data of ATLAS collaboration \cite{ATLAS:2022kqu} in the rapidity range of $-2.5<y<2.0$.
We employ the new Modified Tsallis distribution to describe the transverse flow and in medium energy loss \cite{JPC_Kapil, Khandai}
of hadrons in the medium. Both systematic and statistical uncertainties
are combined and incorporated into the fitting process.

\section{Tsallis/Hagedorn distribution function and the modification}
The transverse mass ($m_{\rm{T}} =\sqrt{p_{\rm{T}}^2+m^2}$) distribution 
of particles produced in hadronic collisions can be described by the
Hagedorn function which is a QCD-inspired summed power
law \cite{Hagedorn:1983wk} given as
\begin{eqnarray}
E~\frac{d^{3}N}{dp^{3}} = A~ \Bigg(1 + \frac{m_{\rm{T}}}{p_{0}}\Bigg)^{-n}~.
\label{Hag}
\end{eqnarray}
This function describes both the bulk spectra in the low $m_{\rm{T}}$ 
region and the particles produced in QCD hard scatterings reflected 
in the high $p_{\rm{T}}$ region. Let us compare this function with the
Tsallis distribution \cite{Tsallis:1987eu, Biro:2008hz} of thermodynamic
origin given by
\begin{eqnarray}
E \frac{d^{3}N}{dp^{3}} = 
C_n ~ m_{\rm{T}} ~ \Bigg(1 + (q-1) \frac{m_{\rm{T}}}{T} \Bigg)^{-1/(q - 1)}~.
\label{Tsallis}
\end{eqnarray}
The Tsallis distribution describes near-thermal systems in terms of
Tsallis parameter $T$ and the parameter $q$ which measures degree of
non-thermalization \cite{q_Tsallis}. The functions in Eq.~\ref{Hag}
and in Eq.~(\ref{Tsallis}) have similar mathematical forms with  
$n = 1/(q - 1)$ and $p_0 = n\,T$. Larger values of $n$ correspond
to smaller values of $q$. Both $n$ and $q$ have been 
interchangeably used in Tsallis distribution
\cite{Biro:2008hz, Adare:2010fe, Cleymans:2012ya, Adare:2011vy, Abelev:2006cs}.
Phenomenological studies suggest that, for quark-quark point 
scattering, $n\sim4$ \cite{Blankenbecler:1975ct, Brodsky:2005fza}, 
which grows larger if multiple scattering centers are involved.
The study in Ref.~\cite{Zheng:2015mhz} suggests that both the 
forms given in Eq.~\ref{Hag} and in Eq.~(\ref{Tsallis}) give equally 
good fit to the hadron spectra in $p+p$ collisions. We use
Eq.~(\ref{Tsallis}) in case of $p+p$ collisions.

Tsallis/Hagedorn function is able to describe $p_{\rm{T}}$ spectra
in $p+p$ collisions practically at all generations of proton colliders.
There have been many attempts to use the Tsallis distribution in
heavy ion collisions as well by including the transverse collective 
flow ~\cite{Tang:2008ud, Khandai:2013fwa, Sett:2015lja}. In addition, 
in heavy ion collisions, particle spectra at high $p_{\rm{T}}$ are 
known to be modified due to in-medium energy loss.
The Tsallis/Hagedorn distribution can be modified by including
these final state effects in different $p_{\rm{T}}$ regions as
follows:

\begin{subequations} \label{modified_new_func_tsallis_distribution_function}
\begin{align} 
E \frac{d^{3}N}{dp^{3}} &= A_{1} \Bigg[\exp\left(-\frac{\beta  p_{\rm{T}}}{p_{1}}\right)
   + \frac{m_{\rm{T}}}{p_{1}}\Bigg]^{-n_{1}} ~ :~  p_{\rm{T}} < p_{\rm{T_{th}}} ~.
\label{new_func_tsallis_distribution_function} \\
E \frac{d^{3}N}{dp^{3}}  &= A_{2}~ \Bigg[\frac{B}{p_{2}}~\Bigg(\frac{p_{\rm{T}}}{q_{0}}\Bigg)^{\alpha}
                              + \frac{m_{\rm{T}}}{p_{2}}\Bigg]^{-n_{2}}~ :~  p_{\rm{T}} > p_{\rm{T_{th}}} ~. 
 \label{new_func_tsallis_distribution_function_second}
\end{align} 
\end{subequations}
The first function (Eq.~\ref{new_func_tsallis_distribution_function}) 
is shown to govern the thermal and collective part of the hadron 
spectrum with the temperature $T=p_{1}/n_{1}$ and the average transverse 
flow velocity $\beta$~\cite{Khandai:2013fwa}.

The second function (Eq.~\ref{new_func_tsallis_distribution_function_second})
is obtained after shifting the distribution in Eq.\ref{Hag} by energy  loss
$\Delta m_{\rm{T}}$ in the medium as 
\begin{equation} 
E \frac{d^{3}N}{dp^{3}} =
A_2~ \Bigg[1 + \frac{m_{\rm{T}} +\Delta m_{\rm{T}}}{p_{2}}\Bigg]^{-n_{2}}~.
\label{tsallis_with_energyloss}
\end{equation}
The energy loss $\Delta m_{\rm{T}}$ is proportional to $p_{\rm{T}}$ at low $p_{\rm{T}}$ 
and in general can be parameterized similar to
the work in Ref.~\cite{Spousta:2016agr} as
\begin{equation}
\Delta m_{\rm{T}} = B ~ \Big(\frac{p_{\rm{T}}}{q_{0}}\Big)^\alpha~. 
\label{spousta_energy_loss}
\end{equation}
 Here, the parameter $\alpha$ quantifies different energy
loss regimes for light quarks in the medium~\cite{Baier:2000mf, De:2011fe}.
 The parameter $B$ is proportional to the medium size and 
$q_{0}$ is an arbitrary scale set as 1~GeV. 
 Using Eq.~(\ref{spousta_energy_loss}) in Eq.~\ref{tsallis_with_energyloss} and
 ignoring 1 we get Eq.~(\ref{new_func_tsallis_distribution_function_second})
 applicable for high  $p_{\rm{T}}$.
 In our study, we find that this function describes the particle spectra
at $p_{\rm{T_{th}}}$ above 7~GeV/$c$. Fits to the data would constrain the
value of $B/p_2$ and thus $p_2$ is not an independent parameter. 
The empirical parton energy loss in nuclear collisions at RHIC 
energies is found to be proportional to $p_{\rm{T}}$ \cite{Wang:2008se}. 
%%%%%%%%%%%%%%%%%%%%%%%%%%%%%%%%%%%%%%%%%%%%%%%%%%%%%%%%%%%%%%%%%%%%%%%%%%%%%
%%%%%%%%%%%%%%%%%%%%%%%%%%%%%%%%%%%%%%%%%%%%%%%%%%%%%%%%%%%%%%%%%%%%%%%%%%%%%

\section{Results and discussions}
%%%%%%%%%%%%%%%%%%%%%%%%%%%%%%%%%%%%%%%%%%%%%%%%%%%%%%%%%%%%%%%%%%%%%%%%%%%%%
%%%%%%%%%%%%%%%%%%%%%%%%%%%%%%%%%%%%%%%%%%%%%%%%%%%%%%%%%%%%%%%%%%%%%%%%%%%%%
\subsection{$p+Pb$ collisions at $\sqrt{s_{\rm{NN}}}$ = 5.02 TeV}
%%%%%%%%%%%%%%%%%%%%%%%%%%%%%%%%%%%%%%%%%%%%%%%%%%%%%%%%%%%%%%%%%%%%%%%%%%%%%
%%%%%%%%%%%%%%%%%%%%%%%%%%%%%%%%%%%%%%%%%%%%%%%%%%%%%%%%%%%%%%%%%%%%%%%%%%%%%

\begin{figure}[!h]
\centering
\includegraphics[width=0.70\linewidth]{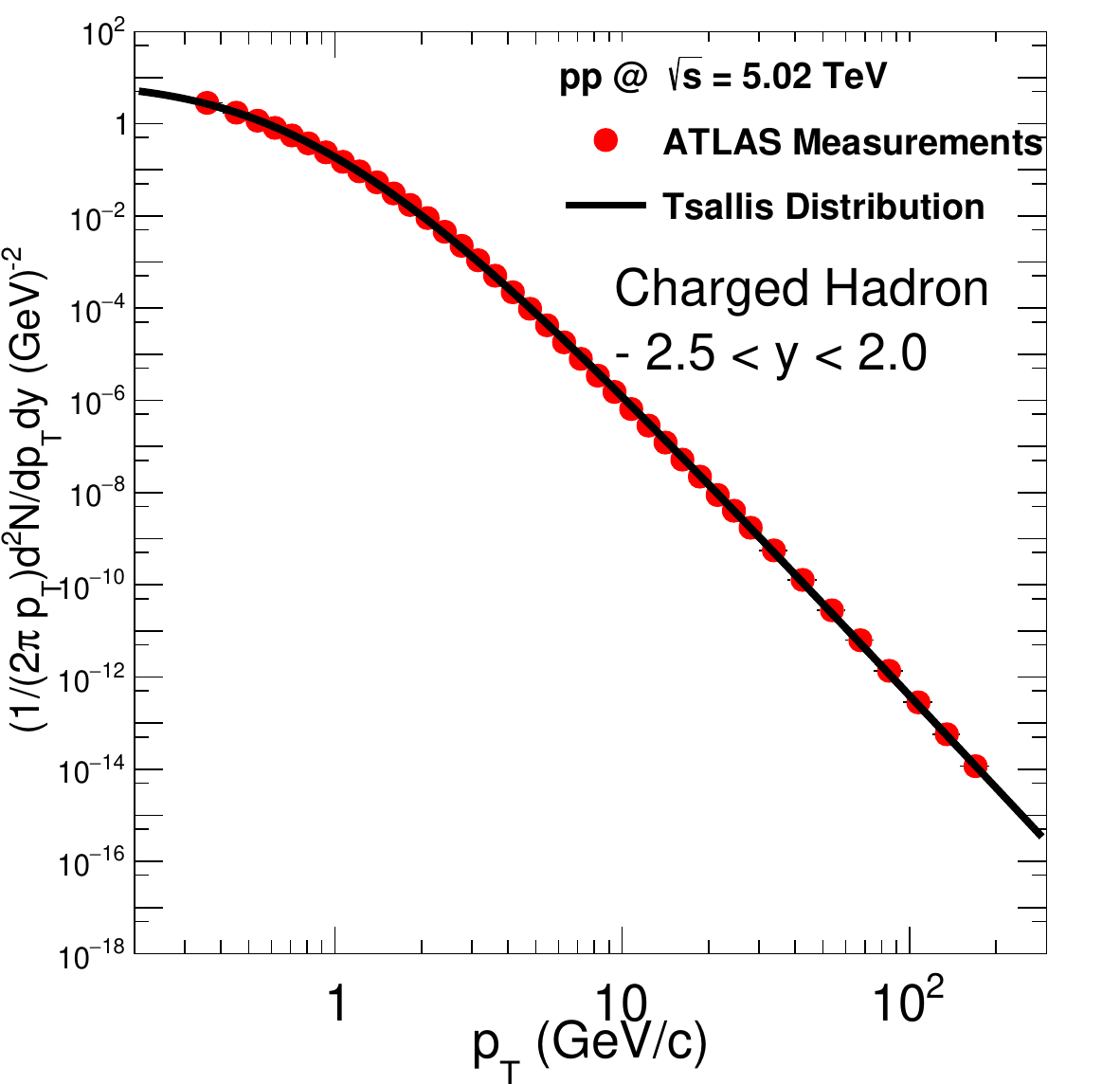}
\caption{The invariant yields of the charged particles as a function of 
transverse momentum $p_{\rm{T}}$ for $p+p$ collision at $\sqrt{s}$ = 5.02 TeV 
measured by the ATLAS experiment  \cite{ATLAS:2022kqu}. The solid curves are
the fitted Tsallis distribution functions.}
\label{Figure1_pp_collision_atlas}
\end{figure}

Figure (\ref{Figure1_pp_collision_atlas}) shows the invariant yields
of the charged
particles as a function of $p_{\rm{T}}$ for $p+p$ collisions at $\sqrt{s}$ = 5.02 TeV
measured by the ATLAS experiment \cite{ATLAS:2022kqu}. The solid curves are the
Tsallis distributions fitted to the spectra. The Tsallis distribution function
gives good description of the data for both the collision energies which
can be inferred from the values  of $\chi^{2}/\rm{NDF}$ given in the
Table~(\ref{Table_one_pp_collision}).

%%%%%%%%%%%%%%%%%%%%%%%%%%%%%%%%%%%%%%%%%%%%%%%%%%%%%%%%%%%%%%%%%%%%%%%%%%%%%
%%%%%%%%%%%%%%%%%%%%%%%%%%%%%%%%%%%%%%%%%%%%%%%%%%%%%%%%%%%%%%%%%%%%%%%%%%%%%
%%%%%%%%%%%%%%%%%%%%%%%%%%%%%%%%%%%%%%%%%%%%%%%%%%%%%%%%%%%%%%%%%%%%%%%%%%%%%

\begin{figure}[htp]
\centering
\includegraphics[width=0.80\linewidth]{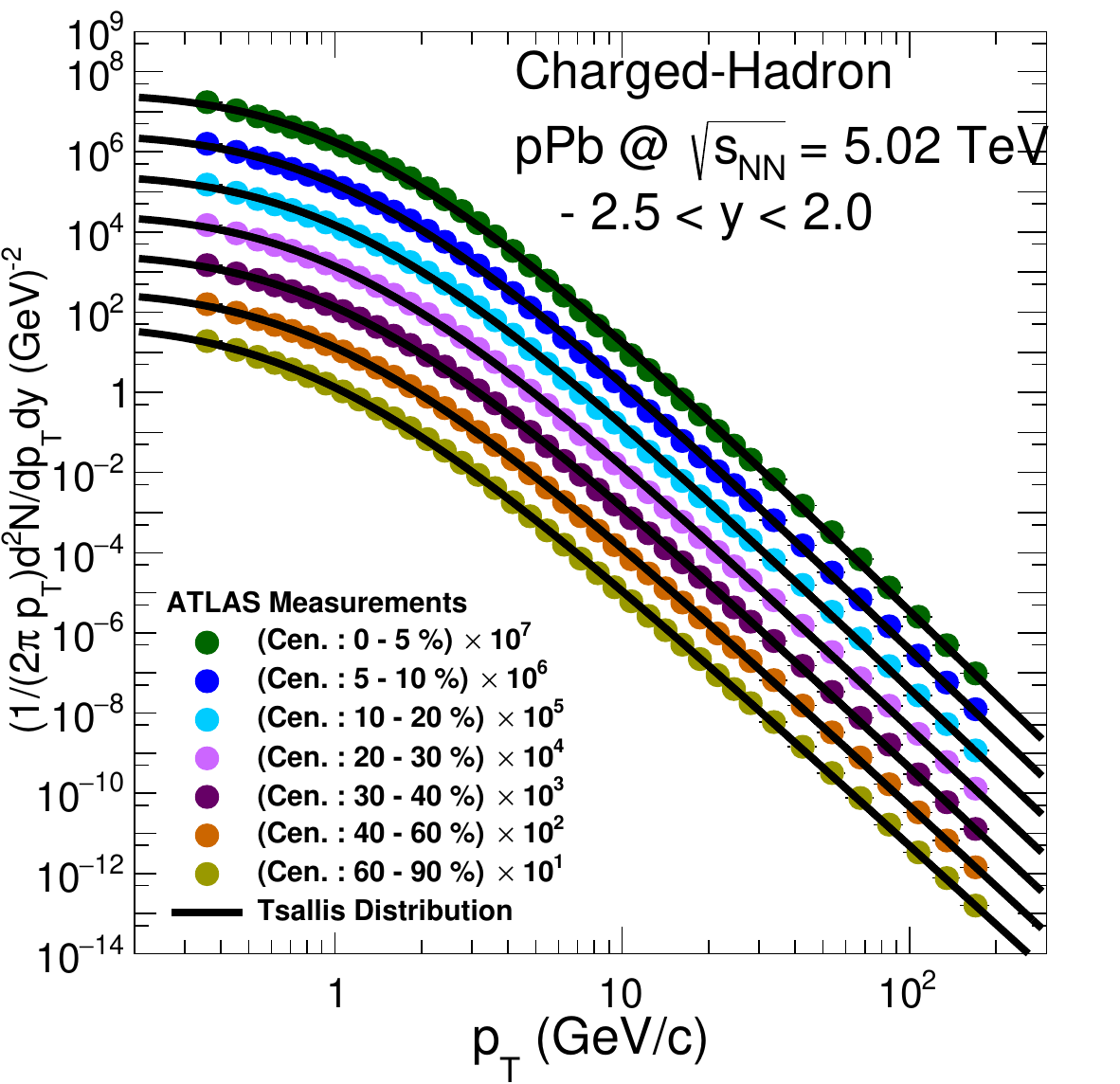}
\caption{The invariant yields of the charged particles  as a function of the  
transverse momentum $p_{\rm{T}}$ for different centrality classes in $p+Pb$ 
collisions at $\sqrt{s_{\rm{NN}}}$ = 5.02 TeV measured by the ATLAS experiment
\cite{ATLAS:2022kqu}.
The solid curves are the fitted Tsallis distribution functions.}
\label{Figure2_pLead_502tev_tsallis}
\end{figure}

\begin{figure}[htp]
\centering
\includegraphics[width=0.80\linewidth]{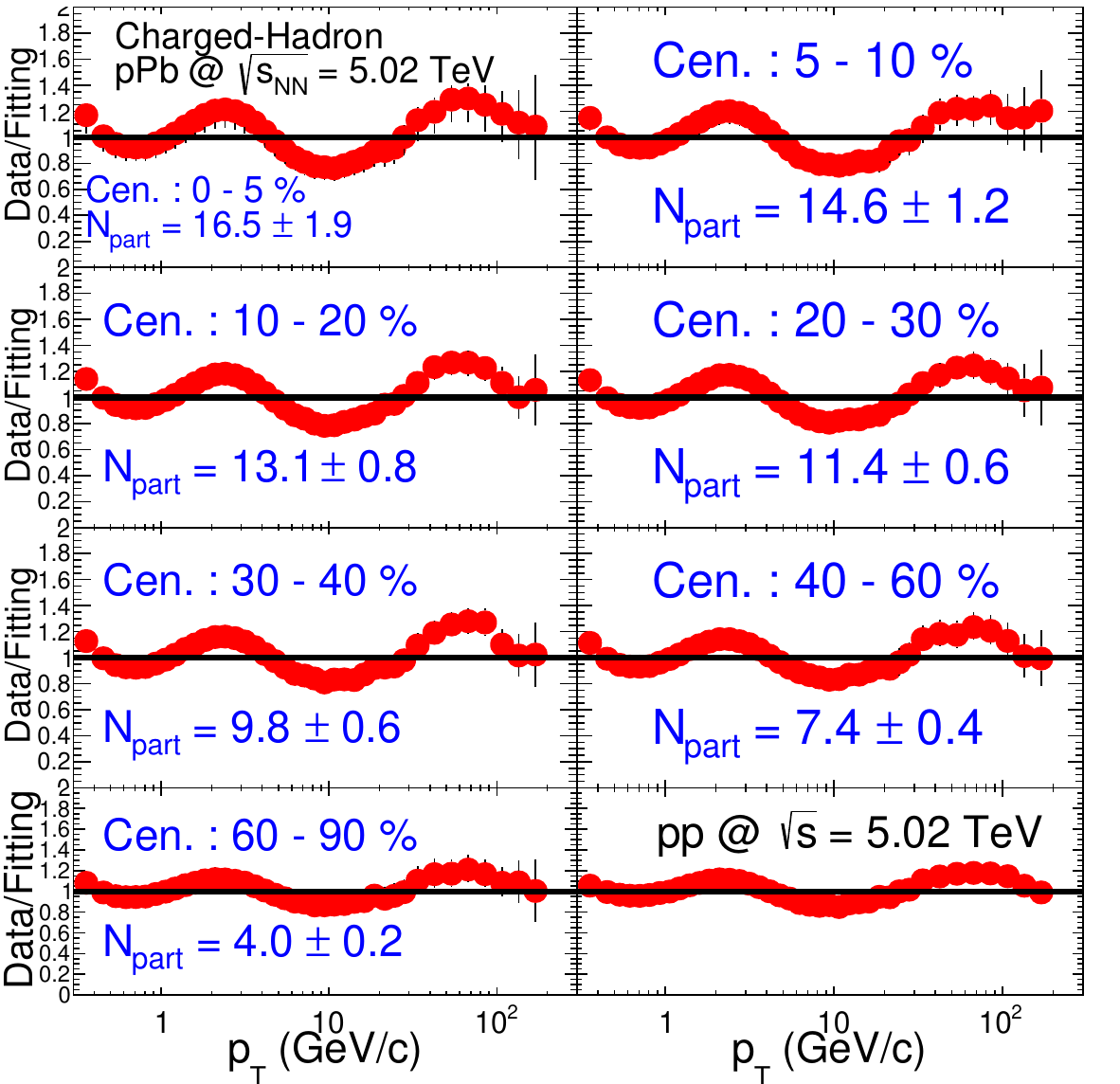}
\caption{The ratio of the charged particles yields data and their Tsallis fits 
as a function of the transverse momentum $p_{\rm{T}}$ for different centrality 
classes in $p+Pb$ collisions at $\sqrt{s_{\rm{NN}}}$ = 5.02 TeV.}
\label{Figure3_pLead_502tev_databyfit}
\end{figure}

%%%%%%%%%%%%%%%%%%%%%%%%%%%%%%%%%%%%%%%%%%%%%%%%%%%%%%%%%%%%%%%%%%%%%%%%%%%%%
%%%%%%%%%%%%%%%%%%%%%%%%%%%%%%%%%%%%%%%%%%%%%%%%%%%%%%%%%%%%%%%%%%%%%%%%%%%%%
%%    Proton - Lead Collisions @ 5.02 TeV :: Tsallis Distribution 
%%%%%%%%%%%%%%%%%%%%%%%%%%%%%%%%%%%%%%%%%%%%%%%%%%%%%%%%%%%%%%%%%%%%%%%%%%%%%
%%%%%%%%%%%%%%%%%%%%%%%%%%%%%%%%%%%%%%%%%%%%%%%%%%%%%%%%%%%%%%%%%%%%%%%%%%%%%
Figure (\ref{Figure2_pLead_502tev_tsallis}) shows the invariant yields
of the charged particles as a function of $p_{\rm{T}}$ for different 
centrality classes in $p+Pb$ collisions at $\sqrt{s_{\rm{NN}}}$ = 5.02 TeV
measured by the ATLAS experiment \cite{ATLAS:2022kqu}.
The solid curves are the fitted Tsallis distributions.
Figure (\ref{Figure3_pLead_502tev_databyfit}) shows the ratio of the data
and the fitted Tsallis distribution as a function of $p_{\rm{T}}$ for $p+Pb$
collisions at $\sqrt{s_{\rm{NN}}}$ = 5.02 TeV. ATLAS measured data of $p+Pb$ and
$p+p$ collisions show deviations from the fit which can be inferred from the
values  of $\chi^{2}/\rm{NDF}$ given in the
Table~(\ref{Table_two_pLead_collision_tsallis}).
%%%%%%%%%%%%%%%%%%%%%%%%%%%%%%%%%%%%%%%%%%%%%%%%%%%%%%%%%%%%%%%%%%%%%%%%%%%%%
%%%%%%%%%%%%%%%%%%%%%%%%%%%%%%%%%%%%%%%%%%%%%%%%%%%%%%%%%%%%%%%%%%%%%%%%%%%%%
%%%%%%%%%%%%%%%%%%%%%%%%%%%%%%%%%%%%%%%%%%%%%%%%%%%%%%%%%%%%%%%%%%%%%%%%%%%%%
%%%%%%%%%%%%%%%%%%%%%%%%%%%%%%%%%%%%%%%%%%%%%%%%%%%%%%%%%%%%%%%%%%%%%%%%%%%%%
%%%%%%%%%%%%%%%%%%%%%%%%%%%%%%%%%%%%%%%%%%%%%%%%%%%%%%%%%%%%%%%%%%%%%%%%%%%%%
%%%%%%%%%%%%%%%%%%%%%%%%%%%%%%%%%%%%%%%%%%%%%%%%%%%%%%%%%%%%%%%%%%%%%%%%%%%%%
%%%%%%%%%%%%%%%%%%%%%%%%%%%%%%%%%%%%%%%%%%%%%%%%%%%%%%%%%%%%%%%%%%%%%%%%%%%%%

\begin{figure}[htp]
\centering
\includegraphics[width=0.80\linewidth]{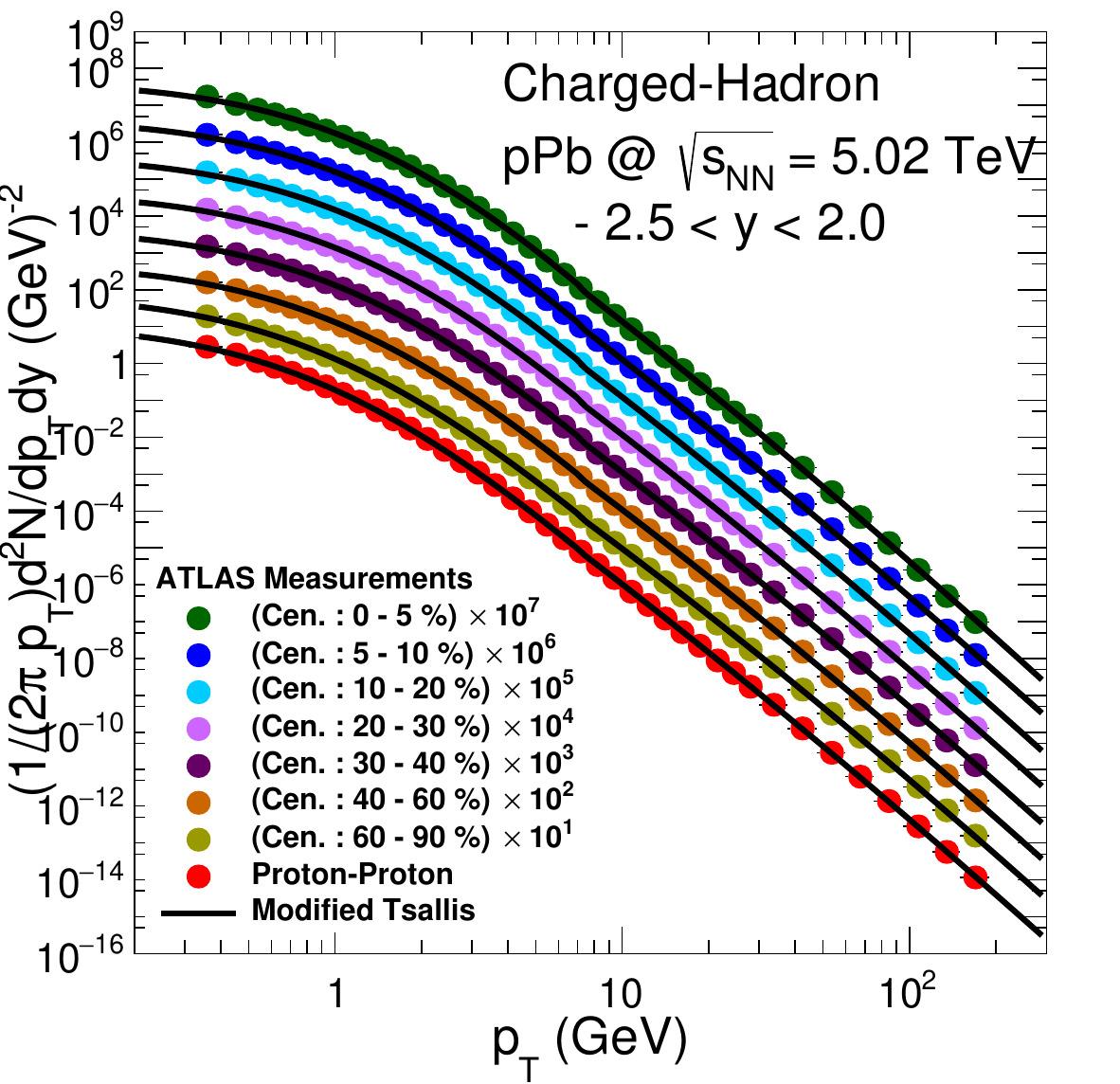}
\caption{The invariant yields of the charged particles  as a function of the  
transverse momentum $p_{\rm{T}}$ for different centrality classes in $p+Pb$
collisions at $\sqrt{s_{\rm{NN}}}$ = 5.02 TeV measured by the ATLAS
experiment \cite{ATLAS:2022kqu}.
The solid curves are the modified Tsallis distributions
(Eq. \ref{new_func_tsallis_distribution_function}).}
\label{Figure4_pLead_502tev_tsallis_modified}
\end{figure}

\begin{figure}[htp]
\centering
\includegraphics[width=0.80\linewidth]{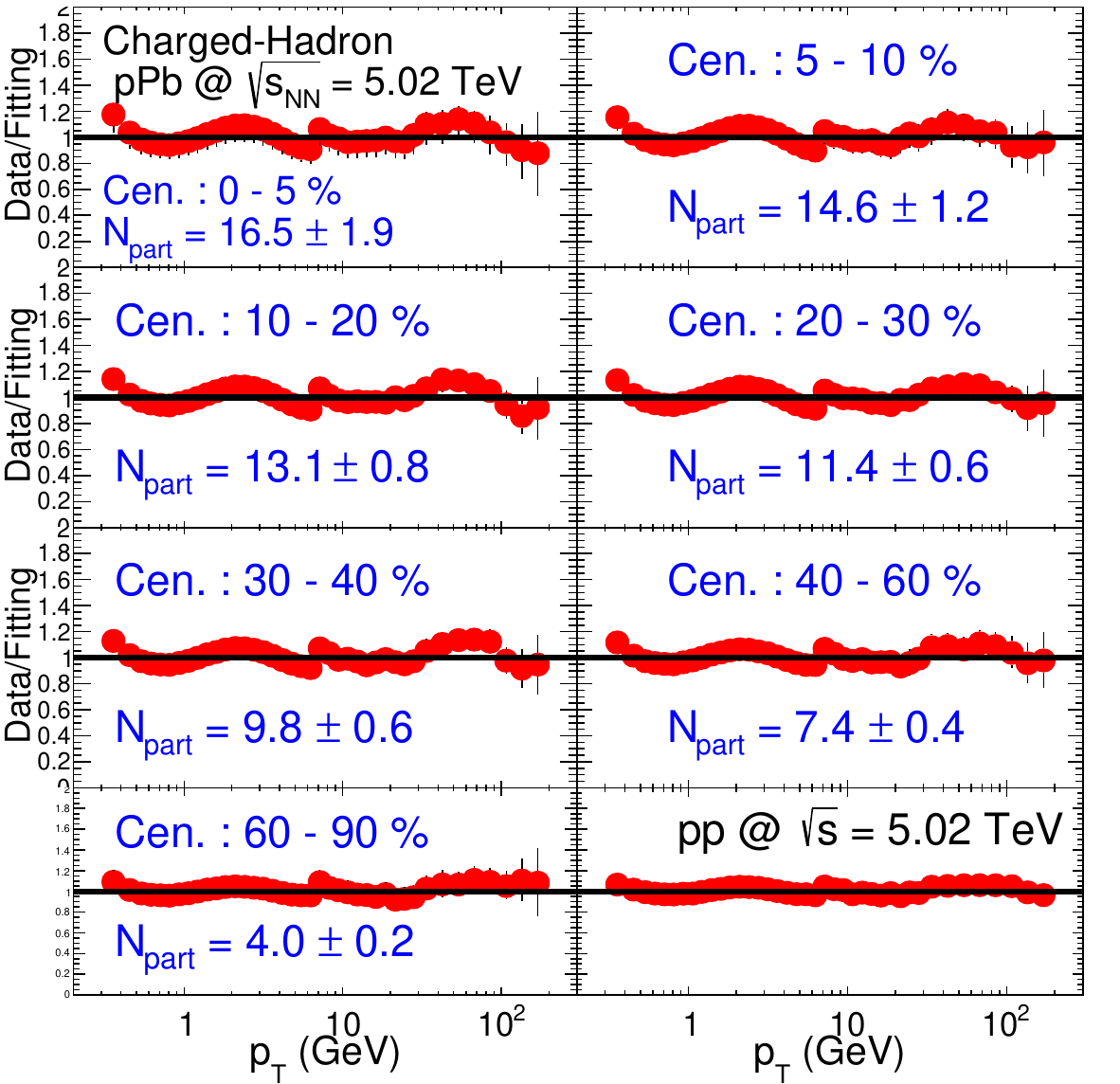}
\caption{The ratio of the charged particle yield data and the fit function  
(Modified Tsallis distribution Eq.~\ref{new_func_tsallis_distribution_function}) 
as a function of the transverse momentum $p_{\rm{T}}$ for different centrality 
classes in $p+Pb$ collisions at $\sqrt{s_{\rm{NN}}}$ = 5.02 TeV.}
\label{Figure5_pLead_502tev_databyfit_modified}
\end{figure}

%%%%%%%%%%%%%%%%%%%%%%%%%%%%%%%%%%%%%%%%%%%%%%%%%%%%%%%%%%%%%%%%%%%%%%%%%%%%%
%%%%%%%%%%%%%%%%%%%%%%%%%%%%%%%%%%%%%%%%%%%%%%%%%%%%%%%%%%%%%%%%%%%%%%%%%%%%%
%%    Proton - Lead Collisions @ 5.02 TeV :: Modified Tsallis Distribution 
%%%%%%%%%%%%%%%%%%%%%%%%%%%%%%%%%%%%%%%%%%%%%%%%%%%%%%%%%%%%%%%%%%%%%%%%%%%%%
%%%%%%%%%%%%%%%%%%%%%%%%%%%%%%%%%%%%%%%%%%%%%%%%%%%%%%%%%%%%%%%%%%%%%%%%%%%%%
Figure (\ref{Figure4_pLead_502tev_tsallis_modified}) shows the invariant  
yields of the charged particles as a function of $p_{\rm{T}}$ for different 
centrality classes in $p+Pb$ collisions at $\sqrt{s_{\rm{NN}}}$ = 5.02 TeV
measured by the ATLAS experiment \cite{ATLAS:2022kqu}.
The solid curves are the modified Tsallis distributions given by
Eq.~(\ref{new_func_tsallis_distribution_function} and 
\ref{new_func_tsallis_distribution_function_second}). 
Figure (\ref{Figure5_pLead_502tev_databyfit_modified}) shows the ratio
of the data and the fit function by the modified Tsallis distribution
as a function of $p_{\rm{T}}$ for different centrality classes in $p+Pb$
collisions at $\sqrt{s_{\rm{NN}}}$ = 5.02 TeV. 
The ratio of the data and the fit function shows that modified Tsallis
distribution function gives excellent description of the measured data
in full $p_{\rm{T}}$ range for all centrality classes.
The parameters of the modified Tsallis distribution are given in the
Table~(\ref{Table_three_pLead_collision_tsallis_modified}).
The values of the first set of parameters ($n_{1}$, $p_{1}$, $\beta$)
are constant for differnt ramge of pseudo-rapidities.
While fitting the second function, we fix the parameter
$n_{2} = 7.67$ guided by $p+p$ value.
The exponent $\alpha$ which decides the variation of the energy 
loss of partons as a function of their energy remains same.
In conclusion, the function given in Eq.~(
\ref{new_func_tsallis_distribution_function} and
\ref{new_func_tsallis_distribution_function_second}) gives excellent
description of the hadron spectra over wide range of $p_{\rm{T}}$
with its parameters indicating different physics effects in the
$p+Pb$ collisions. 
%%%%%%%%%%%%%%%%%%%%%%%%%%%%%%%%%%%%%%%%%%%%%%%%%%%%%%%%%%%%%%%%%%%%%%%%%%%%%
%%%%%%%%%%%%%%%%%%%%%%%%%%%%%%%%%%%%%%%%%%%%%%%%%%%%%%%%%%%%%%%%%%%%%%%%%%%%%
%%%%%%%%%%%%%%%%%%%%%%%%%%%%%%%%%%%%%%%%%%%%%%%%%%%%%%%%%%%%%%%%%%%%%%%%%%%%%
%%%%%%%%%%%%%%%%%%%%%%%%%%%%%%%%%%%%%%%%%%%%%%%%%%%%%%%%%%%%%%%%%%%%%%%%%%%%%
%%%%%%%%%%%%%%%%%%%%%%%%%%%%%%%%%%%%%%%%%%%%%%%%%%%%%%%%%%%%%%%%%%%%%%%%%%%%%
%%%%%%%%%%%%%%%%%%%%%%%%%%%%%%%%%%%%%%%%%%%%%%%%%%%%%%%%%%%%%%%%%%%%%%%%%%%%%

\clearpage
%%%%%%%%%%%%%%%%%%%%%%%%%%%%%%%%%%%%%%%%%%%%%%%%%%%%%%%%%%%%%%%%%%%%%%%%%%%%%
%%%%%%%%%%%%%%%%%%%%%%%%%%%%%%%%%%%%%%%%%%%%%%%%%%%%%%%%%%%%%%%%%%%%%%%%%%%%%
\subsection{$Pb+Pb$ collisions at $\sqrt{s_{\rm{NN}}}$ = 5.02 TeV}
\begin{figure}[htp]
\centering
\includegraphics[width=0.70\linewidth]{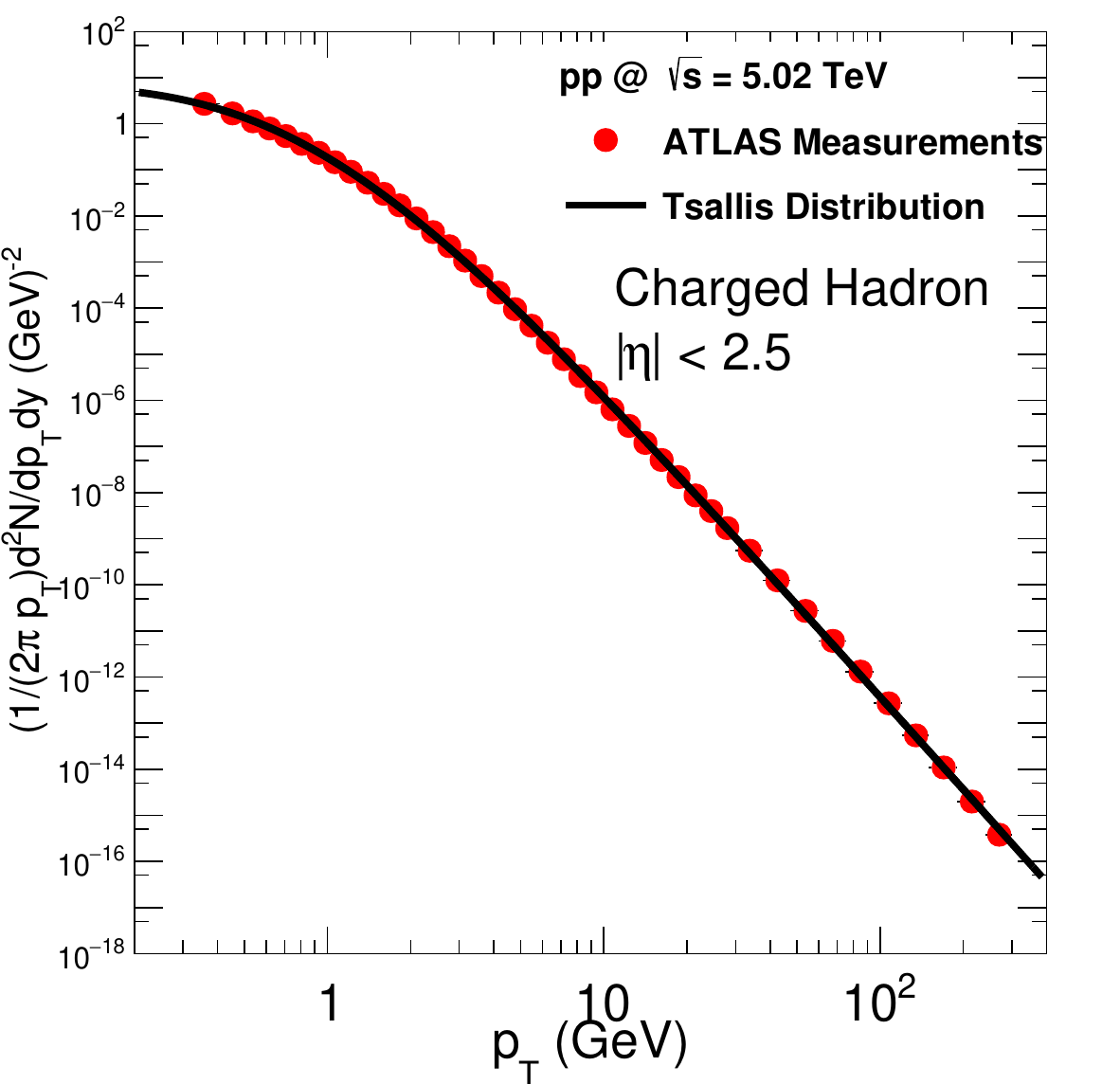}
\caption{The invariant yields of the charged particles as a function of 
transverse momentum $p_{\rm{T}}$ for $p+p$ collision at $\sqrt{s}$ = 5.02 TeV 
measured by the ATLAS experiment  \cite{ATLAS:2022kqu}. The solid curves are
the fitted Tsallis distribution functions.}
\label{Figure6_pp_collision_502tev_refPbPb_atlas}
\end{figure}

%%%%%%%%%%%%%%%%%%%%%%%%%%%%%%%%%%%%%%%%%%%%%%%%%%%%%%%%%%%%%%%%%%%%%%%%%%%%%
%%%%%%%%%%%%%%%%%%%%%%%%%%%%%%%%%%%%%%%%%%%%%%%%%%%%%%%%%%%%%%%%%%%%%%%%%%%%%
%%%%%%%%%%%%%%%%%%%%%%%%%%%%%%%%%%%%%%%%%%%%%%%%%%%%%%%%%%%%%%%%%%%%%%%%%%%%%
%%%%%%%%%%%%%%%%%%%%%%%%%%%%%%%%%%%%%%%%%%%%%%%%%%%%%%%%%%%%%%%%%%%%%%%%%%%%%
%%    Proton - Proton Collisions @ 5.02 TeV :: Tsallis Distribution 
%%%%%%%%%%%%%%%%%%%%%%%%%%%%%%%%%%%%%%%%%%%%%%%%%%%%%%%%%%%%%%%%%%%%%%%%%%%%%
%%%%%%%%%%%%%%%%%%%%%%%%%%%%%%%%%%%%%%%%%%%%%%%%%%%%%%%%%%%%%%%%%%%%%%%%%%%%%
Figure (\ref{Figure6_pp_collision_502tev_refPbPb_atlas}) shows the invariant
yields of the charged particles as a function of $p_{\rm{T}}$ for $p+p$ collisions 
at $\sqrt{s}$ = 5.02 TeV measured by the ATLAS experiment \cite{ATLAS:2022kqu}.
The solid curves are the Tsallis distributions fitted to the spectra. The Tsallis 
distribution function gives good description of the data for both the collision
energies which can be inferred from the values  of $\chi^{2}/\rm{NDF}$ given in
the Table~(\ref{Table_four_pp_collision}).
%%%%%%%%%%%%%%%%%%%%%%%%%%%%%%%%%%%%%%%%%%%%%%%%%%%%%%%%%%%%%%%%%%%%%%%%%%%%%
%%%%%%%%%%%%%%%%%%%%%%%%%%%%%%%%%%%%%%%%%%%%%%%%%%%%%%%%%%%%%%%%%%%%%%%%%%%%%
%%%%%%%%%%%%%%%%%%%%%%%%%%%%%%%%%%%%%%%%%%%%%%%%%%%%%%%%%%%%%%%%%%%%%%%%%%%%%

\begin{figure}[htp]
\centering
\includegraphics[width=0.80\linewidth]{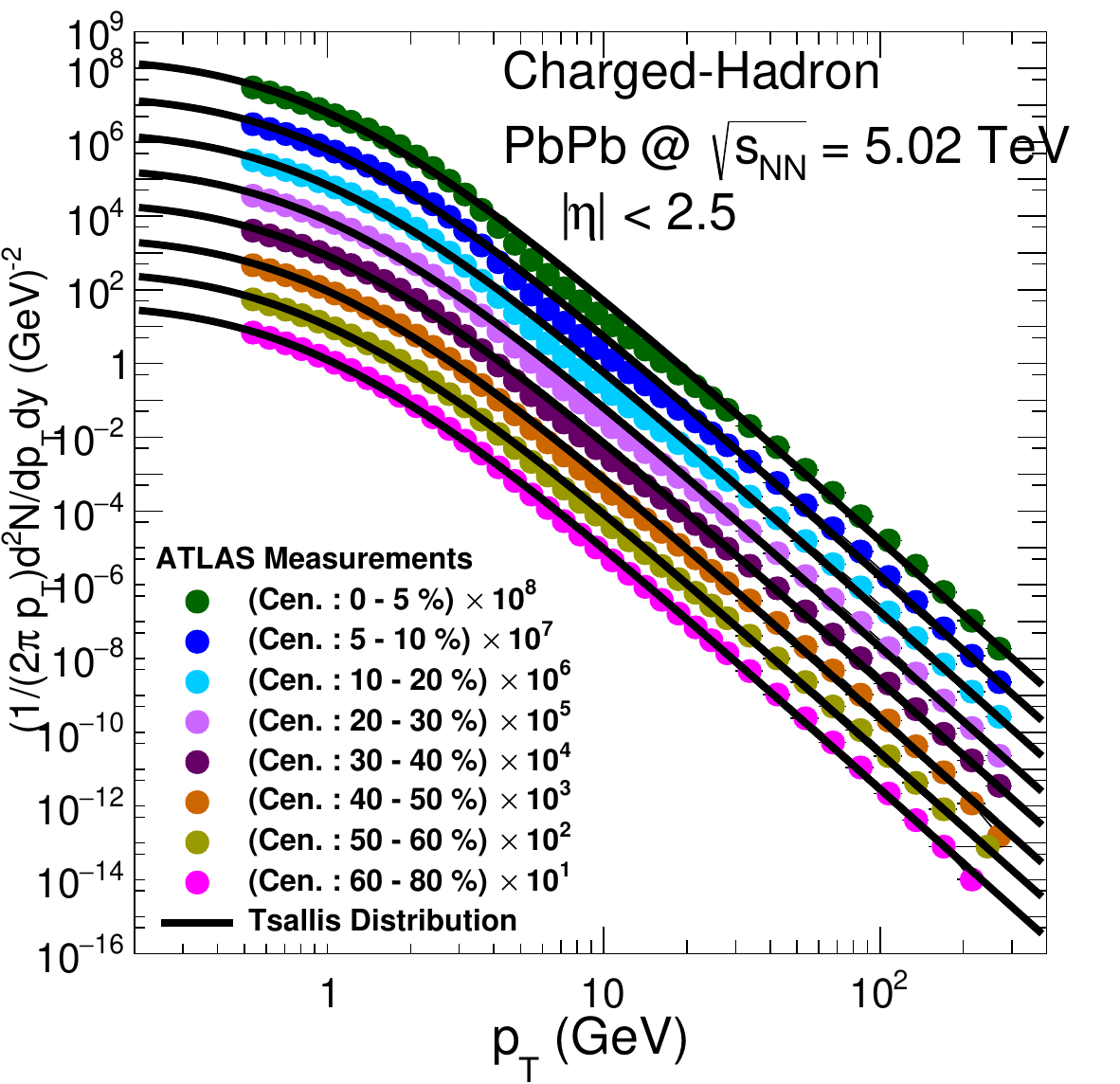}
\caption{The invariant yields of the charged particles  as a function of the  
transverse momentum $p_{\rm{T}}$ for different centrality classes in $Pb+Pb$ 
collisions at $\sqrt{s_{\rm{NN}}}$ = 5.02 TeV measured by the ATLAS experiment
\cite{ATLAS:2022kqu}.
The solid curves are the fitted Tsallis distribution functions.}
\label{Figure7_LeadLead_502tev_tsallis}
\end{figure}

\begin{figure}[htp]
\centering
\includegraphics[width=0.80\linewidth]{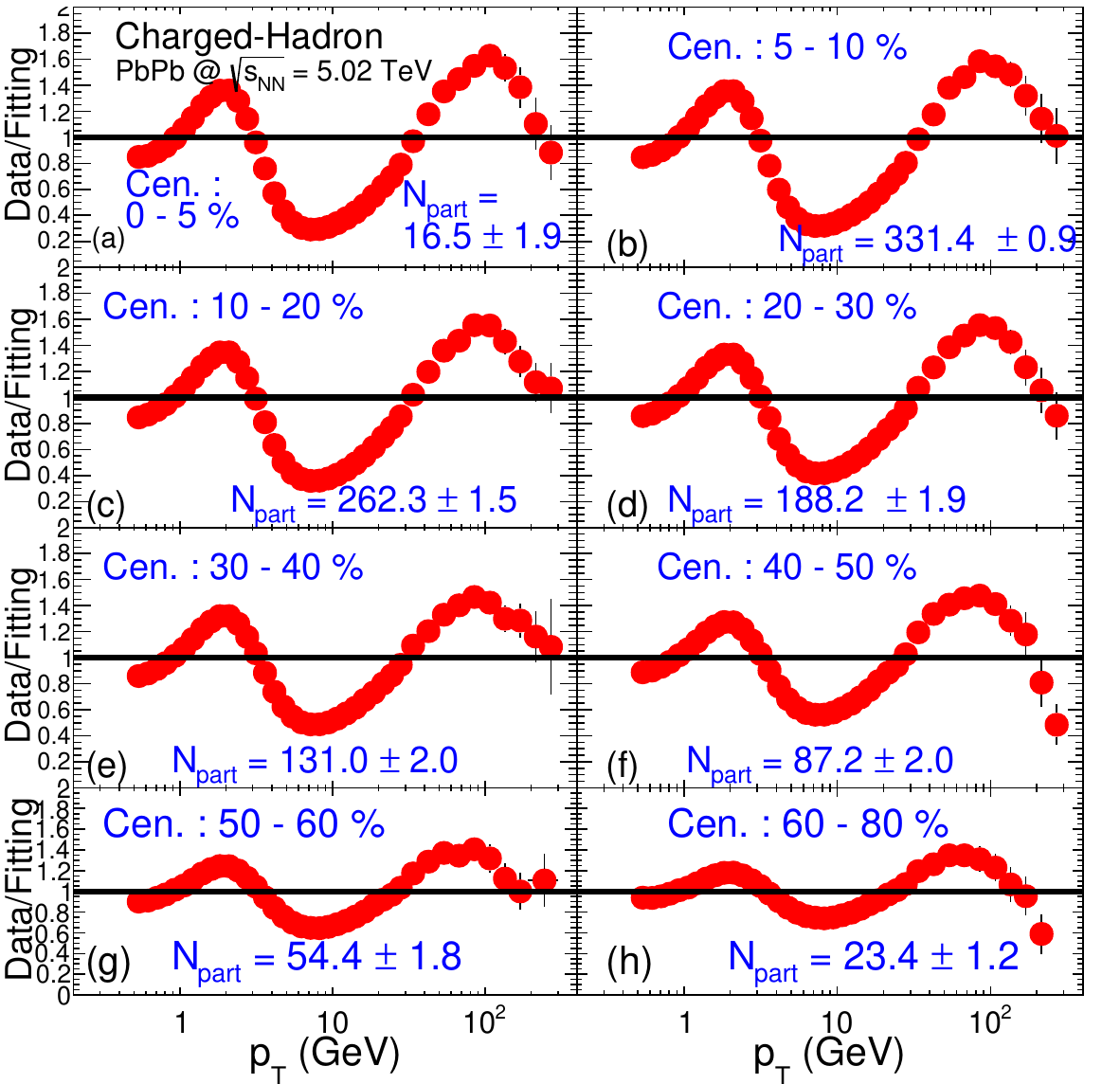}
\caption{The ratio of the charged particles yields data and their Tsallis fits 
as a function of the transverse momentum $p_{\rm{T}}$ for different centrality 
classes in $Pb+Pb$ collisions at $\sqrt{s_{\rm{NN}}}$ = 5.02 TeV.}
\label{Figure8_LeadLead_502tev_databyfit}
\end{figure}

%%%%%%%%%%%%%%%%%%%%%%%%%%%%%%%%%%%%%%%%%%%%%%%%%%%%%%%%%%%%%%%%%%%%%%%%%%%%%
%%%%%%%%%%%%%%%%%%%%%%%%%%%%%%%%%%%%%%%%%%%%%%%%%%%%%%%%%%%%%%%%%%%%%%%%%%%%%
%%    Lead - Lead (PbPb) Collisions @ 5.02 TeV :: Tsallis Distribution 
%%%%%%%%%%%%%%%%%%%%%%%%%%%%%%%%%%%%%%%%%%%%%%%%%%%%%%%%%%%%%%%%%%%%%%%%%%%%%
%%%%%%%%%%%%%%%%%%%%%%%%%%%%%%%%%%%%%%%%%%%%%%%%%%%%%%%%%%%%%%%%%%%%%%%%%%%%%
Figure (\ref{Figure7_LeadLead_502tev_tsallis}) shows the invariant yields
of the charged particles as a function of $p_{\rm{T}}$ for different 
centrality classes in $Pb+Pb$ collisions at $\sqrt{s_{\rm{NN}}}$ = 5.02 TeV
measured by the ATLAS experiment \cite{ATLAS:2022kqu}.
The solid curves are the fitted Tsallis distributions.
Figure (\ref{Figure8_LeadLead_502tev_databyfit}) shows the ratio of the data
and the fitted Tsallis distribution as a function of $p_{\rm{T}}$ for $Pb+Pb$
collisions at $\sqrt{s_{\rm{NN}}}$ = 5.02 TeV. ATLAS measured data of $Pb+Pb$ and
$p+p$ collisions show deviations from the fit which can be inferred from the
values  of $\chi^{2}/\rm{NDF}$ given in the
Table~(\ref{Table_five_pLead_collision_tsallis}).
%%%%%%%%%%%%%%%%%%%%%%%%%%%%%%%%%%%%%%%%%%%%%%%%%%%%%%%%%%%%%%%%%%%%%%%%%%%%%
%%%%%%%%%%%%%%%%%%%%%%%%%%%%%%%%%%%%%%%%%%%%%%%%%%%%%%%%%%%%%%%%%%%%%%%%%%%%%
%%%%%%%%%%%%%%%%%%%%%%%%%%%%%%%%%%%%%%%%%%%%%%%%%%%%%%%%%%%%%%%%%%%%%%%%%%%%%
%%%%%%%%%%%%%%%%%%%%%%%%%%%%%%%%%%%%%%%%%%%%%%%%%%%%%%%%%%%%%%%%%%%%%%%%%%%%%
%%%%%%%%%%%%%%%%%%%%%%%%%%%%%%%%%%%%%%%%%%%%%%%%%%%%%%%%%%%%%%%%%%%%%%%%%%%%%
%%%%%%%%%%%%%%%%%%%%%%%%%%%%%%%%%%%%%%%%%%%%%%%%%%%%%%%%%%%%%%%%%%%%%%%%%%%%%
%%%%%%%%%%%%%%%%%%%%%%%%%%%%%%%%%%%%%%%%%%%%%%%%%%%%%%%%%%%%%%%%%%%%%%%%%%%%%

\begin{figure}[htp]
\centering
\includegraphics[width=0.80\linewidth]{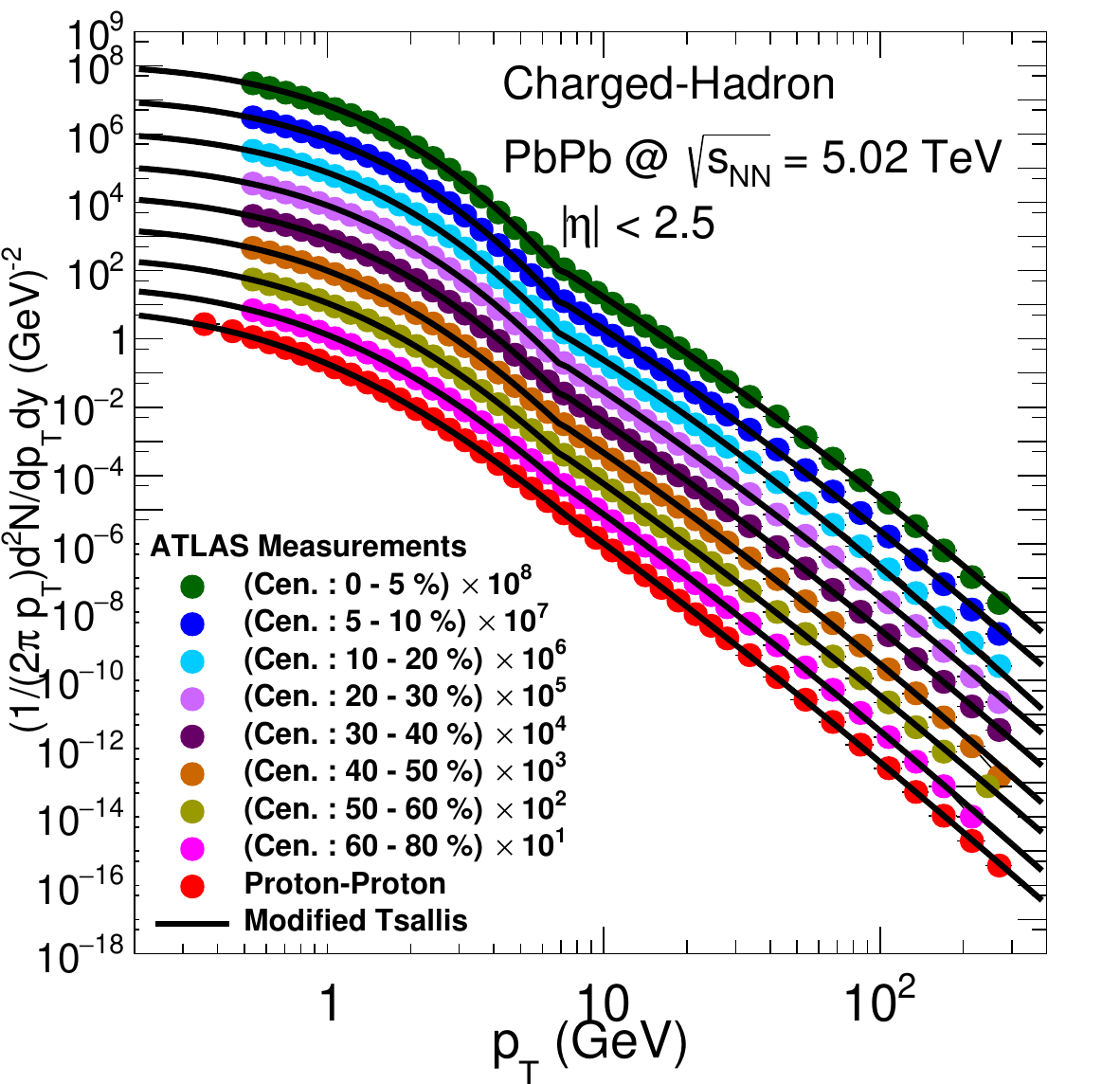}
\caption{The invariant yields of the charged particles  as a function of the  
transverse momentum $p_{\rm{T}}$ for different centrality classes in $Pb+Pb$
collisions at $\sqrt{s_{\rm{NN}}}$ = 5.02 TeV measured by the ATLAS
experiment \cite{ATLAS:2022kqu}.
The solid curves are the modified Tsallis distributions
(Eq. \ref{new_func_tsallis_distribution_function}).}
\label{Figure9_LeadLead_502tev_tsallis_modified}
\end{figure}

\begin{figure}[htp]
\centering
\includegraphics[width=0.80\linewidth]{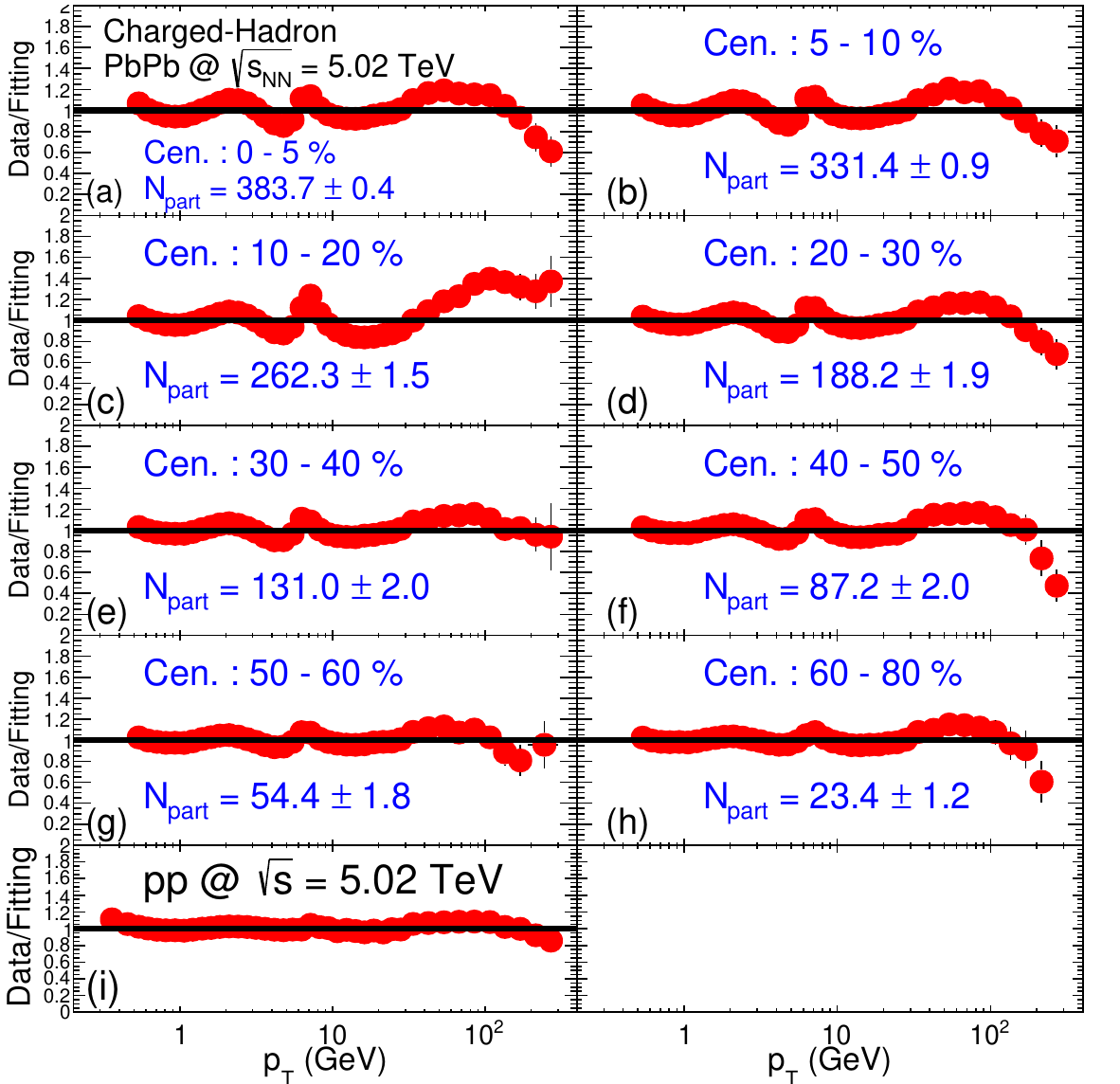}
\caption{The ratio of the charged particle yield data and the fit function  
(Modified Tsallis distribution Eq.~\ref{new_func_tsallis_distribution_function}) 
as a function of the transverse momentum $p_{\rm{T}}$ for different centrality 
classes in $Pb+Pb$ collisions at $\sqrt{s_{\rm{NN}}}$ = 5.02 TeV.}
\label{Figure10_LeadLead_502tev_databyfit_modified}
\end{figure}

%%%%%%%%%%%%%%%%%%%%%%%%%%%%%%%%%%%%%%%%%%%%%%%%%%%%%%%%%%%%%%%%%%%%%%%%%%%%%
%%%%%%%%%%%%%%%%%%%%%%%%%%%%%%%%%%%%%%%%%%%%%%%%%%%%%%%%%%%%%%%%%%%%%%%%%%%%%
%%    Lead - Lead Collisions @ 5.02 TeV :: Modified Tsallis Distribution 
%%%%%%%%%%%%%%%%%%%%%%%%%%%%%%%%%%%%%%%%%%%%%%%%%%%%%%%%%%%%%%%%%%%%%%%%%%%%%
%%%%%%%%%%%%%%%%%%%%%%%%%%%%%%%%%%%%%%%%%%%%%%%%%%%%%%%%%%%%%%%%%%%%%%%%%%%%%
Figure (\ref{Figure9_LeadLead_502tev_tsallis_modified}) shows the invariant  
yields of the charged particles as a function of $p_{\rm{T}}$ for different 
centrality classes in $Pb+Pb$ collisions at $\sqrt{s_{\rm{NN}}}$ = 5.02 TeV measured
by the ATLAS experiment \cite{ATLAS:2022kqu}.
The solid curves are the modified Tsallis distributions given by
Eq.~(\ref{new_func_tsallis_distribution_function} and 
\ref{new_func_tsallis_distribution_function_second}).
Figure (\ref{Figure10_LeadLead_502tev_databyfit_modified}) shows the ratio
of the data and the fit function by the modified Tsallis distribution
as a function of $p_{\rm{T}}$ for different centrality classes in $Pb+Pb$
collisions at $\sqrt{s_{\rm{NN}}}$ = 5.02 TeV.
The ratio of the data and the fit function shows that modified Tsallis
distribution function gives excellent description of the measured data
in full $p_{\rm{T}}$ range for all centrality classes.
The parameters of the modified Tsallis distribution are given in the
Table~(\ref{Table_six_Pb_Pb_collisions_tsallis_modified}).
The values of the first set of parameters ($n_{1}$, $p_{1}$, $\beta$)
are constant for differnt ramge of pseudo-rapidities.
While fitting the second function, we fix the parameter
$n_{2} = 7.70$ guided by $p+p$ value.
The exponent $\alpha$ which decides the variation of the energy 
loss of partons as a function of their energy remains same.
In conclusion, the function given in Eq.~(
\ref{new_func_tsallis_distribution_function} and
\ref{new_func_tsallis_distribution_function_second}) gives excellent
description of the hadron spectra over wide range of $p_{\rm{T}}$
with its parameters indicating different physics effects in the
$Pb+Pb$ collisions. 
%%%%%%%%%%%%%%%%%%%%%%%%%%%%%%%%%%%%%%%%%%%%%%%%%%%%%%%%%%%%%%%%%%%%%%%%%%%%%
%%%%%%%%%%%%%%%%%%%%%%%%%%%%%%%%%%%%%%%%%%%%%%%%%%%%%%%%%%%%%%%%%%%%%%%%%%%%%

\clearpage
%%%%%%%%%%%%%%%%%%%%%%%%%%%%%%%%%%%%%%%%%%%%%%%%%%%%%%%%%%%%%%%%%%%%%%%%%%%%%
%%%%%%%%%%%%%%%%%%%%%%%%%%%%%%%%%%%%%%%%%%%%%%%%%%%%%%%%%%%%%%%%%%%%%%%%%%%%%
\section{Conclusion}
%%%%%%%%%%%%%%%%%%%%%%%%%%%%%%%%%%%%%%%%%%%%%%%%%%%%%%%%%%%%%%%%%%%%%%%%%%%%%
%%%%%%%%%%%%%%%%%%%%%%%%%%%%%%%%%%%%%%%%%%%%%%%%%%%%%%%%%%%%%%%%%%%%%%%%%%%%%

In the article, we carried out an analysis of transverse momentum spectra of
charged hadron in $p+Pb$, and $Pb+Pb$ collisions at $\sqrt{s_{\rm{NN}}}$ = 5.02 TeV.
We first use the Tsallis distribution to describe the $p_{\rm{T}}$ spectra of
charged hardons in  $p+Pb$ and $Pb+Pb$ collisions. We found that Tsallis
distribution does not describe the $p_{\rm{T}}$ spectra properly.
We see a suppression of $p_{\rm{T}}$ spectra above 7GeV/c. To describe and
explain the $p_{\rm{T}}$ spectra of the charged hadrons, we use the modified
Tsallis distribution by incorporating the medium effects.
Here we fit the $p_{\rm{T}}$  spectra of charged hadrons in different
centralities of $p+Pb$ and $Pb+Pb$ collisons at $\sqrt{s_{\rm{NN}}}$ = 5.02 TeV
using modified Tsallis distribution.
We observe the effect of transverse flow in the low-to-intermediate $p_{\rm{T}}$ 
region ($p_{\rm{T}} \leq$ 7.0 GeV/c) and in-medium energy loss in the high
$p_{\rm{T}}$ region ($p_{\rm{T}} >$ 7GeV/c).
We found that in the low (to intermediate) $p_{\rm{T}}$ region the parameters
$n_{1} , p_{1}$ and $\beta$ are more in central collisions and are gradually
decreasing towards peripheral collisions. This is due to the larger number
of multi-scatterings phenomena occurring among partons in the central
collisions than the peripheral collisions.
So there is a transverse collective flow observed among particles in this
region. In the high $p_{\rm{T}}$ region the exponent $\alpha$ which decides
the variation of the energy loss of partons as a function of their energy
remains within 0.59 to 0.73 in $p+Pb$ collisions at $\sqrt{s_{\rm{NN}}}$ =
5.02 TeV, 0.32 to 0.59 in $Pb+Pb$ collisions at $\sqrt{s_{\rm{NN}}}$ = 5.02 TeV.
So, finally, we can say that a simple modification in the Tsallis
distribution gives excellent description of charged particle spectra 
with its parameters having potential to quantify various in-medium effects 
in $p+Pb$ and $Pb+Pb$ collisions.
%%%%%%%%%%%%%%%%%%%%%%%%%%%%%%%%%%%%%%%%%%%%%%%%%%%%%%%%%%%%%%%%%%%%%%%%%%%%%
%%%%%%%%%%%%%%%%%%%%%%%%%%%%%%%%%%%%%%%%%%%%%%%%%%%%%%%%%%%%%%%%%%%%%%%%%%%%%
%%%%%%%%%%%%%%%%%%%%%%%%%%%%%%%%%%%%%%%%%%%%%%%%%%%%%%%%%%%%%%%%%%%%%%%%%%%%%
%%%%%%%%%%%%%%%%%%%%%%%%%%%%%%%%%%%%%%%%%%%%%%%%%%%%%%%%%%%%%%%%%%%%%%%%%%%%%

\clearpage
%%%%%%%%%%%%%%%%%%%%%%%%%%%%%%%%%%%%%%%%%%%%%%%%%%%%%%%%%%%%%%%%%%%%%%%%%%%%%
%%%%%%%%%%%%%%%%%%%%%%%%%%%%%%%%%%%%%%%%%%%%%%%%%%%%%%%%%%%%%%%%%%%%%%%%%%%%%
\section{Appendix :: Table : $p+Pb$ collisions $\sqrt{s_{\rm{NN}}}$ = 5.02 TeV}
%%%%%%%%%%%%%%%%%%%%%%%%%%%%%%%%%%%%%%%%%%%%%%%%%%%%%%%%%%%%%%%%%%%%%%%%%%%%%
%%%%%%%%%%%%%%%%%%%%%%%%%%%%%%%%%%%%%%%%%%%%%%%%%%%%%%%%%%%%%%%%%%%%%%%%%%%%%
%% Table ::  Proton - Proton Collisions @ 5.02 TeV  :: Tsallis Distribution 
%%%%%%%%%%%%%%%%%%%%%%%%%%%%%%%%%%%%%%%%%%%%%%%%%%%%%%%%%%%%%%%%%%%%%%%%%%%%%

\begin{table}[ht]
\caption{The parameters of the Tsallis function obtained by fitting the 
charged particles transverse momentum ($p_{\rm{T}}$) spectrum in $p+p$ collision 
at $\sqrt{s}$ = 5.02 TeV.}
\label{Table_one_pp_collision}
\begin{center}
\begin{tabular}{| c || c | c | c | c | c |} \hline
  Rapidity   &  $n$ & $q$ &   $T$  & $\frac{\chi^{2}}{\rm{NDF}}$ \\ 
  ($y$)   &      &     & (MeV)  &               \\ \hline \hline
- 2.5 $<~ y ~<$ 2.0 & 7.67 $\pm$ 0.10 & 1.13 & 84.78 $\pm$ 5.46 & 0.05 \\ \hline 
\end{tabular}
\end{center}
\end{table}

%%%%%%%%%%%%%%%%%%%%%%%%%%%%%%%%%%%%%%%%%%%%%%%%%%%%%%%%%%%%%%%%%%%%%%%%%%%%%
%%%%%%%%%%%%%%%%%%%%%%%%%%%%%%%%%%%%%%%%%%%%%%%%%%%%%%%%%%%%%%%%%%%%%%%%%%%%%

%%%%%%%%%%%%%%%%%%%%%%%%%%%%%%%%%%%%%%%%%%%%%%%%%%%%%%%%%%%%%%%%%%%%%%%%%%%%%
%%%%%%%%%%%%%%%%%%%%%%%%%%%%%%%%%%%%%%%%%%%%%%%%%%%%%%%%%%%%%%%%%%%%%%%%%%%%%
%% Table ::  Proton - Lead Collisions @ 5.02 TeV  :: Tsallis Distribution 
%%%%%%%%%%%%%%%%%%%%%%%%%%%%%%%%%%%%%%%%%%%%%%%%%%%%%%%%%%%%%%%%%%%%%%%%%%%%%
%%%%%%%%%%%%%%%%%%%%%%%%%%%%%%%%%%%%%%%%%%%%%%%%%%%%%%%%%%%%%%%%%%%%%%%%%%%%%
\begin{table}[ht]
\caption{The parameters of the Tsallis function obtained by fitting the 
charged particles transverse momentum ($p_{\rm{T}}$) spectrum in $p+Pb$ 
collisions $\sqrt{s_{\rm{NN}}}$ = 5.02 TeV.}
\label{Table_two_pLead_collision_tsallis}
\begin{center}
\begin{tabular}{| c || c | c | c | c | c | c |} \hline
  Centrality   & $N_{\rm{part}}$ & $n$ & $q$ &  $T$  & $\frac{\chi^{2}}{\rm{NDF}}$ \\ 
  ($\%$)       &               &     &     & (MeV) &   \\ \hline \hline
0  - 5   & 16.50 $\pm$ 1.90  & 7.97 $\pm$ 0.11  & 1.13  & 116.26 $\pm$ 5.98  & 0.16 \\ \hline 
5  - 10  & 14.60 $\pm$ 1.20  & 7.88 $\pm$ 0.11  & 1.13  & 112.85 $\pm$ 6.14  & 0.15 \\ \hline 
10 - 20  & 13.10 $\pm$ 0.80  & 7.83 $\pm$ 0.10  & 1.13  & 110.56 $\pm$ 5.40  & 0.14 \\ \hline 
20 - 30  & 11.40 $\pm$ 0.60  & 7.77 $\pm$ 0.10  & 1.13  & 107.15 $\pm$ 5.54  & 0.13 \\ \hline 
30 - 40  & 9.80  $\pm$ 0.60  & 7.73 $\pm$ 0.10  & 1.13  & 104.12 $\pm$ 5.56  & 0.11 \\ \hline 
40 - 60  & 7.40  $\pm$ 0.40  & 7.65 $\pm$ 0.10  & 1.13  & 98.44  $\pm$ 5.70  & 0.09 \\ \hline 
60 - 90  & 4.00  $\pm$ 0.20  & 7.53 $\pm$ 0.10  & 1.13  & 86.65  $\pm$ 5.77  & 0.05 \\ \hline 
\end{tabular}
\end{center}
\end{table}
%%%%%%%%%%%%%%%%%%%%%%%%%%%%%%%%%%%%%%%%%%%%%%%%%%%%%%%%%%%%%%%%%%%%%%%%%%%%%
%%%%%%%%%%%%%%%%%%%%%%%%%%%%%%%%%%%%%%%%%%%%%%%%%%%%%%%%%%%%%%%%%%%%%%%%%%%%%

%%%%%%%%%%%%%%%%%%%%%%%%%%%%%%%%%%%%%%%%%%%%%%%%%%%%%%%%%%%%%%%%%%%%%%%%%%%%%
%% Table ::  p-Lead Collisions @ 5.02 TeV  :: Modified Tsallis Distribution 
%%%%%%%%%%%%%%%%%%%%%%%%%%%%%%%%%%%%%%%%%%%%%%%%%%%%%%%%%%%%%%%%%%%%%%%%%%%%%

\begin{table}[ht] 
\caption{The parameters of the modified Tsallis function 
Eq.(~\ref{new_func_tsallis_distribution_function} and 
\ref{new_func_tsallis_distribution_function_second}) 
obtained by fitting the charged particle spectra in $p+Pb$ 
collisions at $\sqrt{s_{\rm{NN}}}$ = 5.02 TeV.}
\label{Table_three_pLead_collision_tsallis_modified}
\begin{center}
%%%%%%%%%%%%%%%%%%%%%%%%%%%%%%%%%%%%%%%%%%%%%%%%%%%%%%%%%%%%%%%%%%%%%%%%%%%%%
\resizebox{1.0\textwidth}{!}{
%%%%%%%%%%%%%%%%%%%%%%%%%%%%%%%%%%%%%%%%%%%%%%%%%%%%%%%%%%%%%%%%%%%%%%%%%%%%%
\begin{tabular}{|c || c | c | c | c | c | c | c | c  |}  \hline
  System & Centrality & $N_{\rm{part}}$ & $n_{1}$ & $p_{1}$    & $\beta$ &  $\alpha$ & $B$         & $\frac{\chi^{2}}{\rm{NDF}}$  \\
         &   ($\%$)  &                &         & (GeV/$c$) &         &            & (GeV/$c$)  & \\ \hline \hline
%%%%%%%%%%%%%%%%%%%%%%%%%%%%%%%%%%%%%%%%%%%%%%%%%%%%%%%%%%%%%%%%%%%%%%%%%%%%%
   $p+Pb$    &   0 - 5            & 16.50 $\pm$ 1.90 & 7.69 $\pm$ 2.21 & 1.38 $\pm$ 1.55 & 0.13 $\pm$ 0.55 & 0.59 $\pm$ 0.27 & 2.40 $\pm$ 0.75 & 0.05 \\ \hline 

   $p+Pb$    &   5 - 10           & 14.60 $\pm$ 1.20 & 7.36 $\pm$ 0.84 & 1.23 $\pm$ 0.28 & 0.15 $\pm$ 0.30 & 0.63 $\pm$ 0.25 & 2.76 $\pm$ 1.35 & 0.05 \\ \hline 

   $p+Pb$    &   10 - 20           & 13.10 $\pm$ 0.80 & 7.32 $\pm$ 1.88 & 1.22 $\pm$ 0.79 & 0.13 $\pm$ 0.12 & 0.59 $\pm$ 0.13 & 2.72 $\pm$ 0.47 & 0.05 \\ \hline 

   $p+Pb$    &   20 - 30           & 11.40 $\pm$ 0.60 & 7.20 $\pm$ 2.19 & 1.16 $\pm$ 0.81 & 0.13 $\pm$ 0.19 & 0.60 $\pm$ 0.05 & 2.90 $\pm$ 0.38 & 0.04 \\ \hline 

   $p+Pb$    &   30 - 40           & 9.80 $\pm$ 0.60 & 7.12 $\pm$ 1.96 & 1.11 $\pm$ 0.68 & 0.13 $\pm$ 0.10 & 0.60 $\pm$ 0.05 & 2.95 $\pm$ 0.38 & 0.04 \\ \hline 

   $p+Pb$    &   40 - 60           & 7.40 $\pm$ 0.40 & 6.99 $\pm$ 2.03 & 1.02 $\pm$ 0.64 & 0.13 $\pm$ 0.09 & 0.59 $\pm$ 0.05 & 3.08 $\pm$ 0.39 & 0.03 \\ \hline 

   $p+Pb$    &   60 - 90           & 4.00 $\pm$ 0.20 & 6.78 $\pm$ 2.30 & 0.86 $\pm$ 0.45 & 0.12 $\pm$ 0.10 & 0.57 $\pm$ 0.04 & 3.26 $\pm$ 0.39 & 0.02 \\ \hline 

   $p+p$    &    -                &  -  & 6.94 $\pm$ 2.36 & 0.86 $\pm$ 1.17 & 0.12 $\pm$ 0.10 & 0.61 $\pm$ 0.05 & 2.92 $\pm$ 0.37 & 0.01 \\ \hline 
%%%%%%%%%%%%%%%%%%%%%%%%%%%%%%%%%%%%%%%%%%%%%%%%%%%%%%%%%%%%%%%%%%%%%%%%%%%%%
\end{tabular}}
\end{center}
\end{table}
%%%%%%%%%%%%%%%%%%%%%%%%%%%%%%%%%%%%%%%%%%%%%%%%%%%%%%%%%%%%%%%%%%%%%%%%%%%%%
%%%%%%%%%%%%%%%%%%%%%%%%%%%%%%%%%%%%%%%%%%%%%%%%%%%%%%%%%%%%%%%%%%%%%%%%%%%%%
%%%%%%%%%%%%%%%%%%%%%%%%%%%%%%%%%%%%%%%%%%%%%%%%%%%%%%%%%%%%%%%%%%%%%%%%%%%%%
%%%%%%%%%%%%%%%%%%%%%%%%%%%%%%%%%%%%%%%%%%%%%%%%%%%%%%%%%%%%%%%%%%%%%%%%%%%%%

\clearpage
%%%%%%%%%%%%%%%%%%%%%%%%%%%%%%%%%%%%%%%%%%%%%%%%%%%%%%%%%%%%%%%%%%%%%%%%%%%%%
%%%%%%%%%%%%%%%%%%%%%%%%%%%%%%%%%%%%%%%%%%%%%%%%%%%%%%%%%%%%%%%%%%%%%%%%%%%%%
\section{Appendix :: Table : $Pb+Pb$ collisions $\sqrt{s_{\rm{NN}}}$ = 5.02 TeV}
%%%%%%%%%%%%%%%%%%%%%%%%%%%%%%%%%%%%%%%%%%%%%%%%%%%%%%%%%%%%%%%%%%%%%%%%%%%%%
%%%%%%%%%%%%%%%%%%%%%%%%%%%%%%%%%%%%%%%%%%%%%%%%%%%%%%%%%%%%%%%%%%%%%%%%%%%%%

%%%%%%%%%%%%%%%%%%%%%%%%%%%%%%%%%%%%%%%%%%%%%%%%%%%%%%%%%%%%%%%%%%%%%%%%%%%%%
%%%%%%%%%%%%%%%%%%%%%%%%%%%%%%%%%%%%%%%%%%%%%%%%%%%%%%%%%%%%%%%%%%%%%%%%%%%%%
%% Table ::  Proton - Proton Collisions @ 5.02 TeV  :: Tsallis Distribution 
%%%%%%%%%%%%%%%%%%%%%%%%%%%%%%%%%%%%%%%%%%%%%%%%%%%%%%%%%%%%%%%%%%%%%%%%%%%%%
\begin{table}[ht]
\caption{The parameters of the Tsallis function obtained by fitting the 
charged particles transverse momentum ($p_{\rm{T}}$) spectrum in $p+p$ collision 
at $\sqrt{s}$ = 5.02 TeV.}
\label{Table_four_pp_collision}
\begin{center}
\begin{tabular}{| c || c | c | c | c | c |} \hline
  Rapidity   &  $n$ & $q$ &   $T$  & $\frac{\chi^{2}}{\rm{NDF}}$ \\ 
  ($y$)   &      &     & (MeV)  &               \\ \hline \hline
 - 2.5 $<~ y ~<$ 2.0   & 7.70 $\pm$ 0.09  & 1.13 & 86.19 $\pm$ 5.31  & 0.05 \\ \hline 
\end{tabular}
\end{center}
\end{table}
%%%%%%%%%%%%%%%%%%%%%%%%%%%%%%%%%%%%%%%%%%%%%%%%%%%%%%%%%%%%%%%%%%%%%%%%%%%%%
%%%%%%%%%%%%%%%%%%%%%%%%%%%%%%%%%%%%%%%%%%%%%%%%%%%%%%%%%%%%%%%%%%%%%%%%%%%%%

%%%%%%%%%%%%%%%%%%%%%%%%%%%%%%%%%%%%%%%%%%%%%%%%%%%%%%%%%%%%%%%%%%%%%%%%%%%%%
%%%%%%%%%%%%%%%%%%%%%%%%%%%%%%%%%%%%%%%%%%%%%%%%%%%%%%%%%%%%%%%%%%%%%%%%%%%%%
%% Table ::  Lead - Lead Collisions @ 5.02 TeV  :: Tsallis Distribution 
%%%%%%%%%%%%%%%%%%%%%%%%%%%%%%%%%%%%%%%%%%%%%%%%%%%%%%%%%%%%%%%%%%%%%%%%%%%%%
%%%%%%%%%%%%%%%%%%%%%%%%%%%%%%%%%%%%%%%%%%%%%%%%%%%%%%%%%%%%%%%%%%%%%%%%%%%%%
\begin{table}[ht]
\caption{The parameters of the Tsallis function obtained by fitting the 
charged particles transverse momentum ($p_{\rm{T}}$) spectrum in $Pb+Pb$ 
collisions $\sqrt{s_{\rm{NN}}}$ = 5.02 TeV.}
\label{Table_five_pLead_collision_tsallis}
\begin{center}
\begin{tabular}{| c || c | c | c | c | c | c |} \hline
  Centrality   & $N_{\rm{part}}$ & $n$ & $q$ &  $T$  & $\frac{\chi^{2}}{\rm{NDF}}$ \\ 
  ($\%$)       &               &     &     & (MeV) &   \\ \hline \hline
  0 - 5    & 383.70 $\pm$ 0.40  & 7.71 $\pm$ 0.07  & 1.13    &   94.07 $\pm$ 5.22  & 1.05 \\ \hline 

 5 - 10   & 331.40 $\pm$ 0.90  & 7.70 $\pm$ 0.10  & 1.13    &   95.28 $\pm$ 8.90  & 0.98 \\ \hline 

 10 - 20   & 262.30 $\pm$  1.50  & 7.70 $\pm$  0.12  & 1.13    &    95.85 $\pm$  11.09  & 0.87 \\ \hline 

 20 - 30   & 188.20 $\pm$  1.90  & 7.71 $\pm$  0.10  & 1.13    &    96.33 $\pm$  8.36  & 0.73 \\ \hline 

 30 - 40   & 131.00 $\pm$  2.00  & 7.67 $\pm$  0.09  & 1.13    &    94.27 $\pm$  7.99  & 0.58 \\ \hline 

 40 - 50   & 87.20 $\pm$  2.00  & 7.71 $\pm$  0.10  & 1.13    &    95.12 $\pm$  7.83  & 0.43 \\ \hline 

 50 - 60   & 54.40 $\pm$  1.80  & 7.68 $\pm$  0.11  & 1.13    &    92.76 $\pm$  7.63  & 0.29 \\ \hline 

 60 - 80   & 23.40 $\pm$  1.20  & 7.72 $\pm$  0.11  & 1.13    &    92.51 $\pm$  7.39  & 0.16 \\ \hline 
\end{tabular}
\end{center}
\end{table}
%%%%%%%%%%%%%%%%%%%%%%%%%%%%%%%%%%%%%%%%%%%%%%%%%%%%%%%%%%%%%%%%%%%%%%%%%%%%%
%%%%%%%%%%%%%%%%%%%%%%%%%%%%%%%%%%%%%%%%%%%%%%%%%%%%%%%%%%%%%%%%%%%%%%%%%%%%%

%%%%%%%%%%%%%%%%%%%%%%%%%%%%%%%%%%%%%%%%%%%%%%%%%%%%%%%%%%%%%%%%%%%%%%%%%%%%%
%% Table :: Lead-Lead Collisions @ 5.02 TeV :: Modified Tsallis Distribution 
%%%%%%%%%%%%%%%%%%%%%%%%%%%%%%%%%%%%%%%%%%%%%%%%%%%%%%%%%%%%%%%%%%%%%%%%%%%%%

\begin{table}[ht] 
\caption{The parameters of the modified Tsallis function 
Eq.(~\ref{new_func_tsallis_distribution_function} and 
\ref{new_func_tsallis_distribution_function_second}) 
obtained by fitting the charged particle spectra in $Pb+Pb$ 
collisions at $\sqrt{s_{\rm{NN}}}$ = 5.02 TeV.}
\label{Table_six_Pb_Pb_collisions_tsallis_modified}
\begin{center}
%%%%%%%%%%%%%%%%%%%%%%%%%%%%%%%%%%%%%%%%%%%%%%%%%%%%%%%%%%%%%%%%%%%%%%%%%%%%%
\resizebox{1.0\textwidth}{!}{  
%%%%%%%%%%%%%%%%%%%%%%%%%%%%%%%%%%%%%%%%%%%%%%%%%%%%%%%%%%%%%%%%%%%%%%%%%%%%%
\begin{tabular}{|c || c | c | c | c | c | c | c | c  |}  \hline
  System & Centrality & $N_{\rm{part}}$ & $n_{1}$ & $p_{1}$    & $\beta$ &  $\alpha$ & $B$         & $\frac{\chi^{2}}{\rm{NDF}}$  \\
         &   ($\%$)  &                &         & (GeV/$c$) &         &            & (GeV/$c$)  & \\ \hline \hline
%%%%%%%%%%%%%%%%%%%%%%%%%%%%%%%%%%%%%%%%%%%%%%%%%%%%%%%%%%%%%%%%%%%%%%%%%%%%%
$Pb+Pb$    &   0 - 5            & 383.70 $\pm$ 0.40 & 10.85 $\pm$ 0.43 & 2.00 $\pm$ 1.77 & 0.28 $\pm$ 0.10 & 0.49 $\pm$ 0.05 & 4.97 $\pm$ 0.57 & 0.06 \\ \hline 

  $Pb+Pb$    &   5 - 10           & 331.40 $\pm$ 0.90 & 10.72 $\pm$ 0.44 & 2.00 $\pm$ 1.06 & 0.27 $\pm$ 0.10 & 0.50 $\pm$ 0.13 & 4.83 $\pm$ 0.71 & 1.29 \\ \hline 

  $Pb+Pb$    &   10 - 20          & 262.30 $\pm$ 1.50 & 10.56 $\pm$ 0.47 & 2.00 $\pm$ 1.03 & 0.25 $\pm$ 0.11 & 0.32 $\pm$ 0.06 & 4.61 $\pm$ 0.67 & 1.28 \\ \hline 

  $Pb+Pb$    &   20 - 30          & 188.20 $\pm$ 1.90 & 10.35 $\pm$ 0.72 & 2.00 $\pm$ 1.23 & 0.21 $\pm$ 0.12 & 0.50 $\pm$ 0.14 & 4.26 $\pm$ 0.60 & 0.07 \\ \hline 

  $Pb+Pb$    &   30 - 40          & 131.00 $\pm$ 2.00 & 10.09 $\pm$ 0.96 & 2.00 $\pm$ 1.41 & 0.15 $\pm$ 0.09 & 0.52 $\pm$ 0.12 & 4.02 $\pm$ 0.55 & 0.02 \\ \hline 

  $Pb+Pb$    &   40 - 50          & 87.20 $\pm$ 2.00 & 8.61 $\pm$ 4.54 & 1.43 $\pm$ 1.27 & 0.24 $\pm$ 0.13 & 0.48 $\pm$ 0.17 & 3.61 $\pm$ 0.49 & 1.16 \\ \hline 

  $Pb+Pb$    &   50 - 60          & 54.40 $\pm$ 1.80 & 7.96 $\pm$ 4.10 & 1.22 $\pm$ 0.83 & 0.24 $\pm$ 0.17 & 0.55 $\pm$ 0.16 & 3.56 $\pm$ 0.77 & 0.02 \\ \hline 

  $Pb+Pb$    &   60 - 80          & 23.40 $\pm$ 1.20 & 7.60 $\pm$ 2.35 & 1.11 $\pm$ 0.85 & 0.19 $\pm$ 0.13 & 0.52 $\pm$ 0.21 & 3.01 $\pm$ 0.52 & 0.02 \\ \hline 

  $p+p$    &     -                &  -  & 7.45 $\pm$ 0.52 & 1.11 $\pm$ 0.74 & 0.02 $\pm$ 0.01 & 0.59 $\pm$ 0.12 & 2.86 $\pm$ 0.65 & 0.01 \\ \hline 
%%%%%%%%%%%%%%%%%%%%%%%%%%%%%%%%%%%%%%%%%%%%%%%%%%%%%%%%%%%%%%%%%%%%%%%%%%%%%
\end{tabular}}
\end{center}
\end{table}
%%%%%%%%%%%%%%%%%%%%%%%%%%%%%%%%%%%%%%%%%%%%%%%%%%%%%%%%%%%%%%%%%%%%%%%%%%%%%
%%%%%%%%%%%%%%%%%%%%%%%%%%%%%%%%%%%%%%%%%%%%%%%%%%%%%%%%%%%%%%%%%%%%%%%%%%%%%
%%%%%%%%%%%%%%%%%%%%%%%%%%%%%%%%%%%%%%%%%%%%%%%%%%%%%%%%%%%%%%%%%%%%%%%%%%%%%
%%%%%%%%%%%%%%%%%%%%%%%%%%%%%%%%%%%%%%%%%%%%%%%%%%%%%%%%%%%%%%%%%%%%%%%%%%%%%

\clearpage 
%%%%%%%%%%%%%%%%%%%%%%%%%%%%%%%%%%%%%%%%%%%%%%%%%%%%%%%%%%%%%%%%%%%%%%%%%%%%%
%%%%%%%%%%%%%%%%%%%%%%%%%%%%%%%%%%%%%%%%%%%%%%%%%%%%%%%%%%%%%%%%%%%%%%%%%%%%%

\end{document}